\documentclass{iopjournal}

\usepackage{bm, amsmath, amsfonts, amssymb, ascmac, mathtools, braket}
\usepackage{multirow}
\usepackage{graphicx}
\usepackage{float, color, xcolor}
\usepackage{comment}
\usepackage{bbm}
\usepackage{tabularx}
\usepackage{graphicx}
\usepackage{float, xcolor}
\usepackage{physics}
\usepackage{comment}
\usepackage{subcaption}
\usepackage{cite}
\usepackage{hyperref}
\hypersetup{
  colorlinks=false,
  pdfborder={0 0 0}
}

\newcommand{\ii}{\text{i}}

\begin{document}

\articletype{Article type} 

\title{Multifractal Analysis of the Non-Hermitian Skin Effect: From Many-Body to Tree Models}

\author{Shu Hamanaka}

\affil{Department of Physics, Kyoto University, Kyoto 606-8502, Japan}

\email{hamanaka.shu.45p@st.kyoto-u.ac.jp}


\begin{abstract}
The non-Hermitian skin effect is an anomalous localization phenomenon induced by nonreciprocal dissipation and has attracted considerable attention in recent years both theoretically and experimentally. 
In this article, we review the multifractal aspects of the non-Hermitian skin effect.
In particular, we discuss how the many-body skin effect exhibits multifractality in many-body Hilbert space, unlike the trivial Hilbert-space occupation of the single-particle skin effect on crystalline lattices. We further highlight that the many-body skin effect can coexist with random-matrix spectral statistics, in contrast to the multifractality associated with many-body localization, which typically accompanies the absence of ergodicity. We also introduce a solvable model on a Cayley tree as an effective description of the many-body Hilbert space, in which the multifractal dimensions can be obtained analytically.
This review provides a unified perspective on multifractal structures in the non-Hermitian skin effect across single-particle, many-body, and tree models, and clarifies their distinctive relation to ergodicity in open quantum systems.
\end{abstract}

\tableofcontents

\section{Introduction}




Multifractality provides a universal description of the complex structures emerging in quantum states between localization and ergodicity~\cite{Halsey-PRA-86}.
In disordered systems, critical single-particle wave functions at the Anderson transition exhibit multifractal behavior~\cite{Evers-RMP,Wegner-80, Castellani-86, Schreiber-91, Mudry-96, Evers-00, Mirlin-00}, reflecting the complex distribution of quantum states.
In the presence of interactions, many-body localized phases also display multifractal statistics due to the intricate structure of the many-body Hilbert space~\cite{Luitz-15, Kravtsov-15,Serbyn-17,Mace-19}, often discussed in the context of the ETH–MBL transition.
More recently, monitored quantum dynamics has also been shown to exhibit multifractal structures~\cite{Zabalo-22, Iaconis-21, Sierant-22}.
In this sense, multifractality provides a unifying language for characterizing complex intermediate structures of quantum states.

The non-Hermitian skin effect provides a distinct mechanism of localization~\cite{Yao-18}. 
In contrast to disorder-induced localization, the skin effect originates from nonreciprocal dissipation, leading to the anomalous accumulation of a macroscopic number of eigenstates at system boundaries. 
This phenomenon has attracted considerable attention due to its intimate relation to non-Hermitian topology~\cite{Rudner-09, Sato-11, Esaki-11, Hu-11, Schomerus-13, Leykam-17, Xu-17, Shen-18, Kozii-17, Gong-18, Kawabata-19,Ma-24,Schindler-23,Nakamura-23,Roccati-24,hamanaka-nonhinh-24,Nakai-25} and its significant impact on open quantum dynamics~\cite{Song-19,Liu-20,Haga-21,Mori-20,kawabata-23,Begg-PRL-24,shen-23,Yoshida-24,Jacopo-24}. 
Experimental realizations have also been reported in a variety of classical and quantum platforms~\cite{Brandenbourger-19,Ghatak-20,ZhangXiu-2021,Helbig-20,Hofmann-20,Xiao-20,Liang-PRL-2022,Gou-PRL-2020}. 

While the skin effect in single-particle systems is now formulated in terms of non-Bloch band theory~\cite{Yokomizo-PRL-2019,Amoeba-24} and point-gap topology~\cite{Zhang-20,Okuma-20}, its extension to interacting many-body systems has emerged as an active area of research, revealing rich phenomena arising from the interplay between many-body interactions and nonreciprocity~\cite{
Mu-PRB-20,Lee-PRB-20,Zhang-PRB-20,Liu-PRB-20,Xu-PRB-20,Lee-PRB-21,ALsallom-PRR-22,Zhang-PRB-22,
Kawabata-PRB-22,Yoshida-PRB-22,Faugo-PRL-22,Begg-PRL-24,Kim-Com-24,Yoshida-24,Jacopo-24, SH-Many,Shen-PRL-24,Shimomura-24,Brighi-PRA-24,Hu-PRL-25,Brighi-PRR-25}. Representative examples include interaction-induced skin effects~\cite{Faugo-PRL-22} and Mott skin effects~\cite{Yoshida-24}. 
Despite these developments, most previous studies have primarily focused on real-space particle number distributions, and a direct quantitative characterization of localization in the many-body Hilbert space has received comparatively less attention. 
In particular, the present article is guided by two closely related questions: \emph{how do many-body skin modes occupy the exponentially large many-body Hilbert space, and which properties of the Hilbert-space structure govern this occupancy?}
As we discuss below, the first question is addressed by the emergence of multifractality in many-body skin modes, while the second is illuminated by the analytically solvable tree model, where the interplay between hierarchical structure and nonreciprocity determines the occupation profile.

In this article, we review the multifractal aspects of the non-Hermitian skin effect across different settings, ranging from single-particle systems to many-body and tree models. 
Specifically, we discuss how the many-body skin effect is characterized by multifractality in many-body Hilbert space, unlike the trivial Hilbert-space occupation of conventional single-particle skin modes on crystalline lattices.
We also highlight a key distinction from many-body localization: this multifractality can coexist with random-matrix spectral statistics.
To shed light on the Hilbert-space structures underlying this multifractal behavior, we also discuss analytically solvable models on tree geometries, where the interplay between hierarchical structure and nonreciprocity allows exact derivations of multifractal dimensions.
Taken together, this review provides a unified perspective on the skin effect from the viewpoint of multifractality, and clarifies its distinctive role in open quantum systems.

The rest of this paper is organized as follows. 
In Sec.~\ref{sec:multi analysis}, we briefly review the basics of multifractal analysis. 
In Sec.~\ref{sec:single}, we apply this analysis to single-particle skin effects and show that conventional skin modes on crystalline lattices do not exhibit multifractality, reflecting their trivial occupation of Hilbert space.
In Sec.~\ref{sec:many}, we discuss how the many-body skin effect exhibits multifractality in many-body Hilbert space, and this multifractality can coexist with random-matrix spectral statistics.
In Sec.~\ref{sec:tree}, we introduce a solvable model on a Cayley tree and present analytical expressions for the multifractal dimensions. 
Section~\ref{sec:conclusion} is devoted to the summary and outlook.

\section{Multifractal analysis}\label{sec:multi analysis}

\subsection{Definition of multifractal dimension}
In this section, we briefly review the basic notions characterizing the multifractal scaling of the wave function.
Let $\ket{\psi}$ be a normalized state with $\mathcal{N}$ components in a chosen basis
$\{\ket{j}\}_{j=1}^{\mathcal{N}}$:
\begin{equation}
\ket{\psi} = \sum_{j=1}^{\mathcal{N}} \psi_j \ket{j}.    
\end{equation}
The multifractal properties of $\ket{\psi}$ are encoded in the scaling behavior of its moments.
Specifically, we consider the generalized inverse participation ratio~\cite{Wegner-80}
\begin{align}
    I_q \coloneqq \sum_{j=1}^{\mathcal{N}} |\psi_j|^{2q},
\end{align}
which scale with system size as
\begin{align}
I_q \propto \mathcal{N}^{-\tau_q}.
\end{align}
Here, $\tau_q$ is a nondecreasing ($\tau_q' \ge 0$) and convex ($\tau_q'' \le 0$) function satisfying
$\tau_0=-1$ and $\tau_1=0$.
The multifractal dimensions $D_q$ defined from the exponents $\tau_q$,
\begin{align}
    \label{eq:Dq}
    D_q \coloneqq \frac{\tau_q}{q-1} = -\frac{1}{q-1} \frac{1}{\ln{\mathcal{N}}} \ln{I_q},
\end{align}
which measure how broadly the wave function occupies the Hilbert space.
Note that $D_1$ is defined by the derivative of $\tau_q$ at $q=1$.

In what follows, we restrict ourselves to $q > 0$, for which the multifractal dimensions satisfy $0 \le D_q \le 1$.
A fully extended state is characterized by $D_q=1$, whereas a state confined to a finite portion of Hilbert space has $D_q=0$.
The intermediate regime $0<D_q<1$ corresponds to states that are extended but not fully delocalized.
When $D_q$ is independent of $q$ in this range, the state is called \emph{monofractal} and is described by a single effective dimension.
By contrast, if $D_q$ varies with $q$, the state is \emph{multifractal}, indicating intricate distribution of wave-function amplitudes across Hilbert space.
In such cases, no single exponent suffices, and an entire family of exponents is required to characterize the state.

The multifractal spectrum $f_\alpha$, defined via the Legendre transformation~\cite{Halsey-PRA-86},
\begin{align}
\label{eq:fa}
    f_\alpha
    \coloneqq
    \alpha q - \tau_q
    \quad
    \text{at } q
    \text{ such that }
    \alpha = \frac{d\tau_q}{dq},
\end{align}
describes the distribution of local scaling exponents $\alpha$, in the sense that the number of components with $|\psi_j|^2 \sim \mathcal{N}^{-\alpha}$ scales as $\mathcal{N}^{f_\alpha}$.
For delocalized, localized, and more generally monofractal states, the spectrum collapses to a single point in the $(\alpha,f_\alpha)$ plane: $(\alpha,f_\alpha)=(1,1)$ for perfectly delocalized states, $(0,0)$ for localized states, and $(D,D)$ for monofractal states with a $q$-independent dimension $D_q=D$.
By contrast, multifractality is characterized by a broadened spectrum with nonzero width, $\alpha_{\rm min}<\alpha<\alpha_{\rm max}$ where $f_\alpha>0$, reflecting the coexistence of multiple scaling structures in the wave function.
Equivalently, such a broadened spectrum corresponds to a $q$-dependent multifractal dimension $D_q$.

\subsection{Multifractality in disordered systems}
Representative examples of different multifractal behaviors are found in disordered systems. 
In the Anderson localized phase, wave functions are localized with $D_q = 0$, whereas in the delocalized phase they satisfy $D_q = 1$. Multifractality, $0 < D_q < 1$ with a $q$-dependent
$D_q$, emerges at the Anderson transition~\cite{Evers-RMP,Wegner-80,Castellani-86,Schreiber-91,Mudry-96,Evers-00} and in many-body localized phases~\cite{Luitz-15,Kravtsov-15,Serbyn-17,Mace-19,Monteiro-PRR-21,De-PRB-21}.
Multifractality has also been observed in monitored quantum dynamics~\cite{Zabalo-22,Iaconis-21,Sierant-22}.

In the following, we investigate the multifractal dimensions associated with the non-Hermitian skin effect.
In particular, we contrast conventional single-particle skin modes on crystalline lattices, which do not exhibit multifractality, with the many-body skin effect, where multifractal scaling emerges.
We also introduce an analytically solvable model on a tree geometry, which provides further insight into the emergence of multifractal behavior from the underlying Hilbert space structure.

\section{Single-particle skin effect}\label{sec:single}

\subsection{Hatano-Nelson model}
The non-Hermitian skin effect is an anomalous localization phenomenon induced by nonreciprocal dissipation.
The Hatano-Nelson model~\cite{Hatano-PRL-1996, Hatano-PRB-1997} in the absence of disorder provides the simplest setting for illustrating this effect. The Hamiltonian is given by
\begin{equation}
    \label{eq:single-HN}
    H = \sum_{j} 
    \Big( 
    t_{\rm R} \ket{j+1}\bra{j} 
    + 
    t_{\rm L} \ket{j}\bra{j+1} 
    \Big),
\end{equation}
where $\{\ket{j}\}$ ($j=1,\ldots,L$) denotes the single-particle basis, and $t_{\rm R}$ ($t_{\rm L}$) is the right (left) hopping amplitude.
Below, we analyze the multifractal dimension of eigenstates under both periodic and open boundary conditions.

\subsection{Periodic boundary conditions}

Under periodic boundary conditions, the right eigenstates are plane waves
\begin{equation}
\psi_j^{(n)} = \frac{1}{\sqrt{L}} e^{\ii k_n j},
\end{equation}
with $k_n = 2\pi n/L$ ($n=0,1,\ldots,L-1$).
Then we can directly compute multifractal dimensions as
\begin{equation}
    D_q = -\frac{1}{q-1}\frac{1}{\ln L} \ln \Big(\sum_{j=1}^L \abs{\frac{e^{\ii k_n j}}{\sqrt{L}}}^{2q} \Big)= 1
\end{equation}
Hence, the multifractal dimension is unity, corresponding to a delocalized state.

\subsection{Open boundary conditions}
Under open boundary conditions, the right eigenstates take the form
\begin{equation}
\psi_j^{(n)}
\propto
\beta^j
\sin(k_n j),
\qquad
\beta \coloneqq \sqrt{\frac{t_{\rm R}}{t_{\rm L}}},
\end{equation}
with $k_n = \pi n/(L+1)$ ($n=1,\ldots,L$)
(see Ref.~\cite{Yokomizo-PRL-2019}).
These eigenstates localize at the right (left) edge for $\beta>1$ ($\beta<1$), giving rise to the non-Hermitian skin effect.
For analytical simplicity, we approximate the eigenstates as
\begin{equation}
\psi_j^{(n)}
\simeq
\frac{1}{\sqrt{N}}
\left(
\beta e^{\ii k_n}
\right)^j,
\end{equation}
with the normalization constant
\begin{equation}
N
=
\sum_{j=1}^{L}
\beta^{2j}
=
\frac{\beta^2(\beta^{2L}-1)}{\beta^2-1}.
\end{equation}
This approximation is consistent with the non-Bloch band theory~\cite{Yao-18,Yokomizo-PRL-2019} and captures the essential nature of the skin effect.
Then we can directly evaluate the multifractal dimensions as
\begin{align}
    D_q 
    &\simeq -\frac{1}{q-1}\frac{1}{\ln{L}}  \ln \left( \sum_{i=1}^{L} \left| \frac{\left( \beta e^{\ii k_n} \right)^i}{\sqrt{N}} \right|^{2q} \right) 
    = -\frac{1}{q-1} \frac{1}{\ln{L}} \ln \left( \left( \frac{\beta^2 - 1}{\beta^2 \left( \beta^{2L} - 1 \right)} \right)^q \frac{\beta^{2q} \left( \beta^{2qL} -1 \right)}{\beta^{2q} - 1} \right) \nonumber \\
    &\simeq -\frac{1}{q-1} \frac{1}{\ln L} \ln \left( \frac{\left( \beta^2 - 1 \right)^q}{\beta^{2q} - 1} \right) \rightarrow 0,
\end{align}
where we have taken the limit $L\to\infty$ with $\beta>1$.
Hence, the multifractal dimension vanishes, corresponding to the localized state.

Beyond one dimension, conventional crystalline skin modes are typically localized or, at most, monofractal rather than multifractal.
For a state on a $d$-dimensional lattice that is exponentially localized along $d-n$ directions but extended along the remaining $n$ directions, one finds $D_q=n/d$, which is independent of $q$.
Thus, skin effects in higher dimensions, including higher-order skin effect~\cite{Okugawa-20,Kawabata-PRB-20} exhibit only $q$-independent monofractal scaling rather than multifractality.

\section{Multifractality of many-body non-Hermitian skin effect}\label{sec:many}
In the previous section, we found that single-particle skin modes are perfectly localized and therefore have vanishing multifractal dimensions. We now turn to the many-body skin effect, where multifractal behavior emerges~\cite{SH-Many}. We first introduce the model and then examine its multifractal scaling. We also discuss how such multifractal behavior can coexist with random-matrix spectral statistics (Table~\ref{tab1}).

\begin{table}[H]
\caption{Spectral statistics and multifractality for various quantum phases of matter.}
\centering
\begin{tabular}{ccc}
\hline
Quantum phase & Spectral statistics & Multifractality \\
\hline
Thermalization & Random matrix & Ergodic ($D_q=1$)  \\
Many-body localization & Poisson & Multifractal ($0<D_q<1$) \\
Non-Hermitian skin effect & Random matrix & Multifractal ($0<D_q<1$) \\
\hline
\end{tabular}
\label{tab1}
\end{table}

\subsection{Model}
To investigate the generic properties of the many-body non-Hermitian skin effect that do not rely on special integrable structures or model-specific details, we consider the following nonintegrable non-Hermitian spin chain:
\begin{align}
    H &= \sum_{i=1}^{L} \left[
    \frac{t}{2} \left(
    \left( 1+\gamma \right) \sigma_{i}^{-} \sigma_{i+1}^{+}
    + \left( 1-\gamma \right) \sigma_{i}^{+} \sigma_{i+1}^{-}
    \right)
    + J \sigma_{i}^{z} \sigma_{i+1}^{z}
    + g \sigma_{i}^{x}
    + h \sigma_{i}^{z}
    \right],
    \label{eq: Ham}
\end{align}
where $t, \gamma, J, g, h \in \mathbb{R}$.
Here, $\sigma_i^{x}$, $\sigma_i^{y}$, and $\sigma_i^{z}$ denote Pauli matrices at site $i$, and
$\sigma_i^{\pm} = \sigma_i^{x} \pm \ii \sigma_i^{y}$ are spin raising and lowering operators.
The parameter $\gamma$ quantifies the nonreciprocity in the magnetization hopping. In the Hermitian limit $\gamma=0$, the model reduces to the XXZ chain in the presence of both longitudinal and transverse fields, a standard example of a nonintegrable many-body system in the thermal phase~\cite{Kim-Huse-15}. The nonreciprocal XX coupling considered here is also closely connected to the asymmetric simple exclusion process~\cite{Gwa-Spohn-92L,Gwa-Spohn-92A,Kim-95}.
In the limit $J=g=h=0$, the model reduces essentially to the Hatano-Nelson model~\eqref{eq:single-HN}, for which the single-particle skin modes, as discussed above, have a trivial multifractal structure with vanishing multifractal dimensions. As we show below, this stands in marked contrast to the many-body skin effect, which exhibits multifractal behavior.

\subsection{Multifractal scaling}
\begin{figure}[h]
\centering
\includegraphics[width=\linewidth]{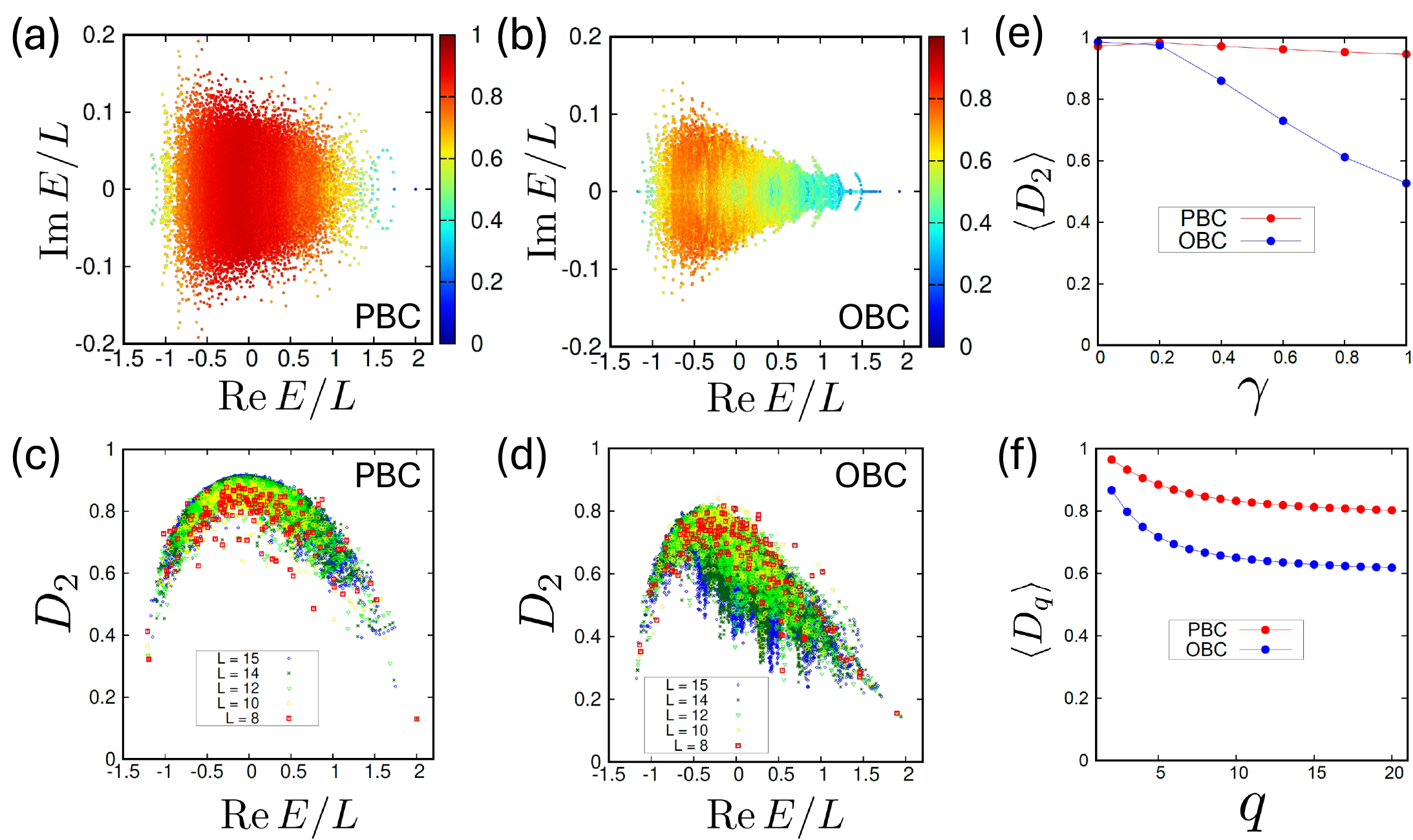}
\caption{Multifractal scaling of the non-Hermitian spin chain in Eq.~\eqref{eq: Ham} ($t=1/\sqrt{2}$, $J=1$, $g=\left( 5+\sqrt{5}\right)/8$, $h = \left( 1+\sqrt{5}\right)/4$).
(a, b)~Eigenvalues $\left( \mathrm{Re}\,E/L, \mathrm{Im}\,E/L \right)$ scaled by the system length $L=15$ under the (a)~periodic boundary conditions (PBC) and (b)~open boundary conditions (OBC) ($\gamma = 0.8$).
The color bars show the multifractal dimension $D_2$ for each right eigenstate.
(c, d)~Multifractal dimensions $D_2$ of individual right eigenstates as a function of $\mathrm{Re}\,E/L$ for the different system lengths $L$ under (c)~PBC and (d)~OBC ($\gamma = 0.8$).
(e)~Multifractal dimension $\langle{D_2}\rangle$ averaged over all right eigenstates as functions of non-Hermiticity $\gamma$ under both PBC (red dots) and OBC (blue dots).
(f)~$q$-dependence of average multifractal dimension $\langle{D_q}\rangle$ under both PBC (red dots) and OBC (blue dots) ($\gamma=0.4$). 
Figure~\ref{fig: Hamiltonian} adapted with permission from Ref.~\cite{SH-Many}. \href{https://doi.org/10.1103/PhysRevB.111.035144} {Copyright (2025) by the American Physical Society.}
}
    \label{fig: Hamiltonian}
\end{figure}
We now examine the multifractal scaling of the non-Hermitian spin chain in Eq.~\eqref{eq: Ham}.
Throughout this section, we use the spin configuration basis as the computational basis.
The parameters are chosen to ensure nonintegrability in the Hermitian limit~\cite{Kim-Huse-15}.
By exact diagonalization, we obtain the complex eigenvalues and the corresponding right eigenstates under both periodic and open boundary conditions.

Figure~\ref{fig: Hamiltonian} (a, b) shows the complex spectrum together with the multifractal dimension $D_2$ for individual right eigenstates.
The spectrum changes significantly between periodic and open boundary conditions, signaling the skin effect.
Note that the symmetry of the spectrum about the real axis originates from the time-reversal symmetry of the Hamiltonian, $H^{*} = H$.
Under periodic boundary conditions, most eigenstates have $D_2 \simeq 1$, whereas under open boundary conditions, their counterparts exhibit substantially smaller multifractal dimensions.
To further resolve the energy dependence, Fig.~\ref{fig: Hamiltonian} (c, d) plots $D_2$ as a function of the real part of the energy for different system lengths.
Under periodic boundary conditions, the distribution of multifractal dimensions forms a characteristic arc-like structure, reminiscent of that found in nonintegrable Hermitian many-body systems~\cite{Backer-19}.
In particular, eigenstates in the middle of the spectrum tend toward $D_2 \simeq 1$ with increasing system length, while those near the spectral edges remain smaller.
This deviation from unity near the spectral edges is naturally attributed to locality constraints of the Hamiltonian, in contrast to fully random matrices, where the corresponding distribution is flat~\cite{Backer-19}.
By contrast, under open boundary conditions, multifractal dimensions remain at intermediate values even in the spectral bulk, indicating that the skin effect suppresses the effective occupation of the many-body Hilbert space.

To extract the multifractal dimension in the thermodynamic limits, we analyze the finite-size scaling of the participation entropy $S_q \coloneqq \ln I_q/(1-q)$, which is fitted as $\langle S_q\rangle = \langle D_q\rangle \ln \mathcal{N} + \langle c_q\rangle$ by averaging over all right eigenstates.
Figure~\ref{fig: Hamiltonian} (e) shows the resulting average multifractal dimension $\langle D_2 \rangle$ as a function of the nonreciprocity $\gamma$.
While $\langle D_2 \rangle$ remains close to unity under periodic boundary conditions, it decreases with increasing $\gamma$ under open boundary conditions, indicating that the skin effect enforces a progressively smaller occupation of the many-body Hilbert space.
We also find a clear $q$ dependence of the multifractal dimension [Fig.~\ref{fig: Hamiltonian} (f)].
Taken together, these results show that the many-body skin effect exhibits multifractality in many-body Hilbert space, as opposed to the vanishing multifractal dimension of the single-particle skin effect.
More detailed analyses of the parameter dependence, together with the corresponding results for integrable systems, can be found in Ref.~\cite{SH-Many}.

\subsection{Spectral statistics}

\begin{figure}[t]
\centering
\includegraphics[width=0.7\linewidth]{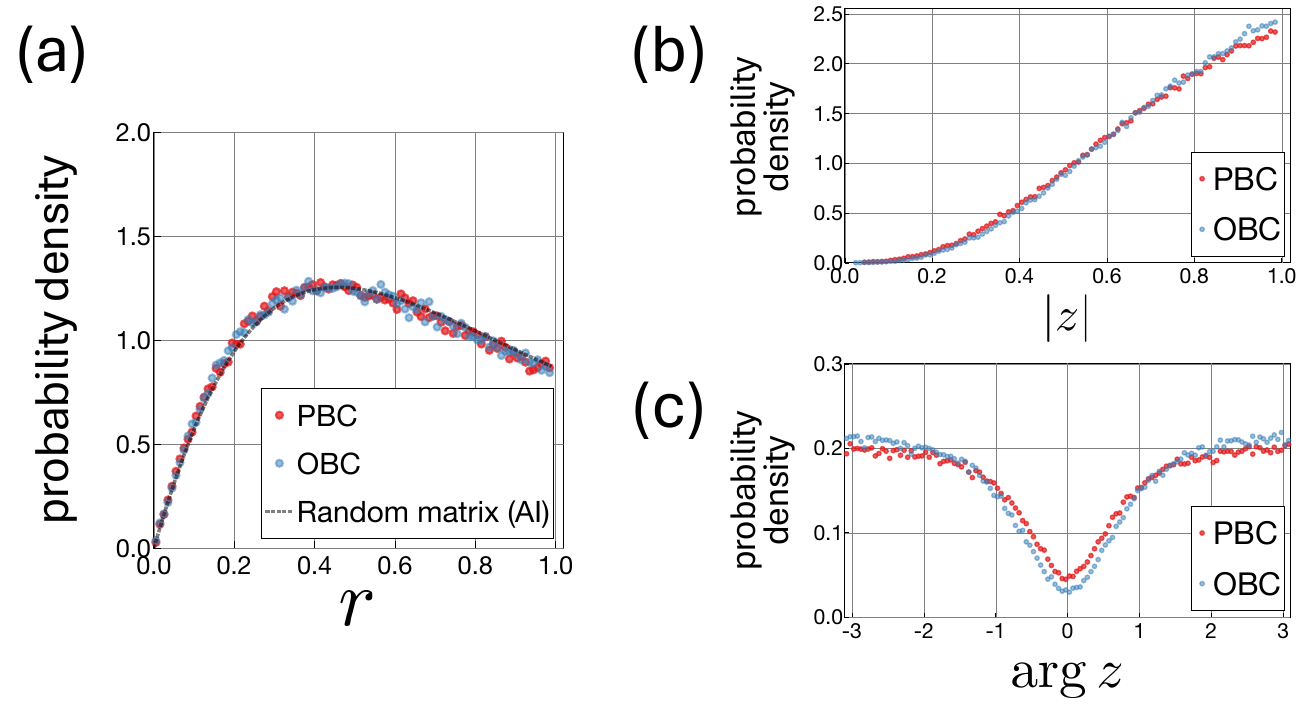}
\caption{
Level-spacing-ratio statistics of the non-Hermitian spin chain in Eq.~\eqref{eq: Ham} under PBC (red dots) and OBC (blue dots) ($t = 1/\sqrt{2}$, $\gamma = 0.6$, $J=1$, $g = (5+\sqrt{5})/8$, $h=(1+\sqrt{5})/4$, $L=14$).
All data are taken away from the spectral edges and the symmetry line, and are averaged over $50$ disorder realizations.
(a) Spacing-ratio distribution $r$ of singular values.
The averages are $\langle r \rangle = 0.5297$ for PBC and $\langle r \rangle = 0.5299$ for OBC.
The black dashed curve shows the analytical result for small non-Hermitian random matrices in class AI, namely $p(r)=27(r+r^2)/[4(1+r+r^2)^{5/2}]$, with $\langle r \rangle = 4-2\sqrt{3} \simeq 0.5359$~\cite{Kawabata-23SVD}.
(b, c) Level-spacing-ratio statistics of complex eigenvalues, shown for (b) the absolute value $|z|$ and (c) the argument $\arg z$.
The averages are $\langle |z| \rangle = 0.7275$ and $\langle \cos \arg z \rangle = -0.1842$ for PBC, and $\langle |z| \rangle = 0.7365$ and $\langle \cos \arg z \rangle = -0.2355$ for OBC.
For comparison, $10^4 \times 10^4$ non-Hermitian random matrices yield $\langle |z| \rangle = 0.7381$ and $\langle \cos \arg z \rangle = -0.2405$~\cite{Sa-20}. Figure~\ref{fig: SVD} adapted with permission from Ref.~\cite{SH-Many}.  \href{https://doi.org/10.1103/PhysRevB.111.035144} {Copyright (2025) by the American Physical Society.}}
\label{fig: SVD}
\end{figure}

In the previous subsection, we saw that the many-body skin effect gives rise to multifractal structure in many-body Hilbert space.
In conventional multifractal quantum phases, such as many-body localized phases, multifractality is typically accompanied by Poisson spectral statistics, reflecting the absence of level repulsion due to local integrals of motion~\cite{Huse-review,Abanin-review}.
By contrast, the multifractality of the many-body skin effect can coexist with random-matrix spectral statistics. 
This coexistence suggests a novel dissipative quantum-chaotic regime distinct from more conventional multifractal phases.
To probe its quantum-chaotic nature, we examine both the singular-value statistics and the complex level-spacing statistics of the non-Hermitian Hamiltonian in Eq.~\eqref{eq: Ham}.
We note that spectral diagnostics of chaos have recently been extensively employed in open quantum systems~\cite{Xu-19, Hamazaki-19, Denisov-19, Can-19JPhysA, Hamazaki-20, Akemann-19, Sa-20, JiachenLi-21, GarciaGarcia-22PRX, GJ-23, Sa-23, Kawabata-23SYK,Roccati-PRB-24}.

We first consider the spacing-ratio statistics of singular values.
For an ordered set of singular values $\{s_n\}$ ($n=1,2,\cdots,\mathcal{N}$), we define the ratio~\cite{Kawabata-23SVD}
\begin{equation}
    r_n \coloneqq \min \left(
    \frac{s_{n+1}-s_n}{s_n-s_{n-1}},
    \frac{s_n-s_{n-1}}{s_{n+1}-s_n}
    \right),
    \qquad
    0 \le r_n \le 1.
\end{equation}
To remove unwanted symmetries, we add a weak disordered term
$\sum_{i=1}^{L} \varepsilon_i \sigma_i^z \sigma_{i+1}^z$,
where each $\varepsilon_i$ is uniformly distributed in $[-0.1,0.1]$.
As shown in Fig.~\ref{fig: SVD} (a), the singular-value statistics under both periodic and open boundary conditions are in close agreement with the random-matrix prediction.
In particular, the averaged spacing ratio takes $\langle r \rangle = 0.5297$ for periodic boundary conditions and $\langle r \rangle = 0.5299$ for open boundary conditions, both close to the random-matrix value $\langle r \rangle = 0.5359$.

We next analyze the statistics of complex level-spacing ratios~\cite{Sa-20},
\begin{equation}
    z_n \coloneqq \frac{E_n^{\rm NN}-E_n}{E_n^{\rm NNN}-E_n},
\end{equation}
where $E_n^{\rm NN}$ ($E_n^{\rm NNN}$) is the nearest (next-nearest) eigenvalue to $E_n$ in the complex-energy plane.
The numerical distributions of both $|z|$ and $\arg z$ are likewise consistent with non-Hermitian random-matrix statistics, as shown in Fig.~\ref{fig: SVD} (b, c).
The small deviation observed under periodic boundary conditions is naturally attributed to the residual effect of average translation symmetry.
Specifically, we obtain $\langle |z| \rangle = 0.7275$ and $\langle \cos \arg z \rangle = -0.1842$ for periodic boundary conditions, and $\langle |z| \rangle = 0.7365$ and $\langle \cos \arg z \rangle = -0.2355$ for open boundary conditions, in good agreement with the corresponding random-matrix values $\langle |z| \rangle = 0.7381$ and $\langle \cos \arg z \rangle = -0.2405$.

Taken together, these results highlight a characteristic feature of the many-body skin effect: the coexistence of multifractal eigenstates and random-matrix spectral statistics, unlike many-body localized phases (Table~\ref{tab1}).

\section{Tree geometry and analytic multifractality}\label{sec:tree}
In this section, we turn to a non-Hermitian model on the Cayley tree~\cite{SH-Cayley}.
We begin by explaining the motivation for introducing the tree model in the present context, and then define the model.
We subsequently construct the relevant basis states and derive the exact eigenstates.
Finally, we compute the multifractal dimensions and discuss the degeneracy of eigenstates.

\subsection{Motivation}
The previous section numerically uncovered the multifractal structure of the many-body skin effect.
While this analysis demonstrates the emergence of multifractality, the relation between the resulting occupation profile and the underlying Hilbert-space structure is not yet fully transparent.
To gain analytical insight, it is therefore useful to consider a simpler setting in which these structural effects can be examined more transparently.

A key starting point is to view many-body Hilbert space as a graph.
In this picture, each spin configuration basis state corresponds to a node, and the Hamiltonian connects nearby nodes through its off-diagonal matrix elements.
Because the number of connected basis states grows hierarchically, the resulting graph is often locally tree-like in character.
Such graph-based descriptions were originally discussed in the context of interacting quantum dots~\cite{Altshuler-PRL-97}, and have since been employed more broadly in studies of multifractality in many-body systems~\cite{Tikonov-review-21}.

Motivated by this spirit, we now turn to a non-Hermitian model on a Cayley tree as an analytically tractable toy model for the many-body Hilbert space.
Although the Cayley tree does not capture the full complexity of the many-body problem, it retains an essential feature of many-body Hilbert space, namely its hierarchical branching structure.
As shown below, the competition between the hierarchical growth of Hilbert space and nonreciprocity determines the multifractal dimensions exactly and clarifies how the occupation profile of skin modes is shaped by the underlying Hilbert-space structure.

\subsection{Model}

\begin{figure}[h]
\centering
\includegraphics[width=\linewidth]{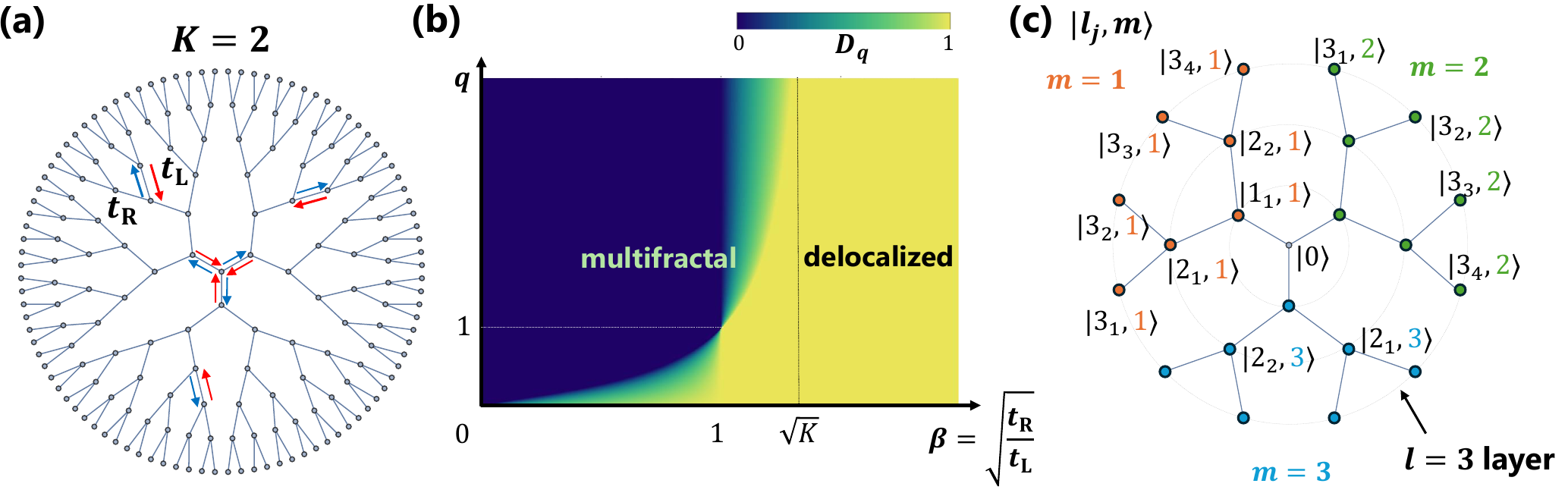}
\caption{(a) Sketch of the model with nonreciprocal hopping on the Cayley tree for branching number $K=2$. The number of layers is $M=6$.
The hopping amplitude from the boundary to the center (from the center to the boundary) is $t_{\rm L}$ ($t_{\rm R}$).(b) Phase diagram of the multifractal dimension $D_q$. For $K\ge 2$, the symmetric eigenstates are localized
in the limit $\beta\to0$, delocalized for $\beta>\sqrt{K}$, and multifractal in the intermediate regime $0<\beta<\sqrt{K}$. (c) Sketch of the symmetric basis construction in the case $K=2$. The symmetric basis states are constructed by linear superposition of the position basis $\ket{l,j, m}$. Figure~\ref{fig:graph} adapted with permission from Ref.~\cite{SH-Cayley}. \href{https://doi.org/10.1103/PhysRevB.111.075162} {Copyright (2025) by the American Physical Society.}}
    \label{fig:graph}
\end{figure}

We study a non-Hermitian model defined on a Cayley tree.
A Cayley tree is a rooted graph characterized by a fixed branching number $K$: apart from the outermost sites, each node connects to one node in the preceding layer and to $K$ nodes in the following layer, so that each bulk node has $K+1$ neighbors [Fig.~\ref{fig:graph} (a)].
The graph is built recursively from a central root node.
Starting from this root, one first attaches $K+1$ nodes to form the first layer.
Each node in a given layer then generates $K$ descendants in the next layer.
After repeating this construction up to depth $M$, one obtains an $M$-layer Cayley tree with branching number $K$.
By construction, all nodes within the same layer are at equal graph distance from the root.
Figure~\ref{fig:graph} (a) shows the case with $M=6$ and $K=2$.

To realize the non-Hermitian skin effect on the Cayley tree, we consider the following nonreciprocal Hamiltonian:
\begin{equation}\label{eq:hamiltonian}
    H =
    t_{\rm R} \sum_{\langle i > j \rangle} \ket{i}\bra{j}
    +
    t_{\rm L} \sum_{\langle i < j \rangle} \ket{i}\bra{j}, 
\end{equation}
where $t_{\rm R}, t_{\rm L} > 0$ denote the hopping amplitudes.
Here, $\sum_{\langle i > j \rangle}$ \big($\sum_{\langle i < j \rangle}$\big) denotes the sum over neighboring nodes $i$ and $j$ such that node $j$ lies closer to (farther from) the central node than node $i$ [see Fig.~\ref{fig:graph} (a)].
Nonreciprocity is introduced when $\beta \neq 1$ ($\beta \coloneqq \sqrt{t_{\rm R}/t_{\rm L}})$, or equivalently when $t_{\rm R} \neq t_{\rm L}$.
For $t_{\rm R} < t_{\rm L}$ ($t_{\rm R} > t_{\rm L}$), the asymmetric hopping biases the particle motion toward the central node (the surface nodes). 
A non-Hermitian tree model was also studied independently~\cite{sun-PRB-24}, while another independent study considered a different context~\cite{Hatano-arxiv-2024}.

\subsection{Symmetric sector}
The full Hilbert-space dimension for an $M$-layer Cayley tree with branching number $K$ in Eq.~\eqref{eq:hamiltonian} is
\begin{equation}
    \mathcal{N}
    = 1 + (K+1)\sum_{l=1}^{M} K^{l-1}
    = 1 + (K+1)\frac{K^M-1}{K-1}.
\end{equation}
While the Hamiltonian in Eq.~\eqref{eq:hamiltonian} is exactly solved~\cite{SH-Cayley}, much of the following discussion is devoted to its symmetric sector for simplicity.

The symmetric sector is spanned by basis states constructed by uniformly superposing nodes within each layer of a given branch, as explained in detail below.
We choose the central node as one of the symmetric basis states as $  |0) \coloneqq \ket{0}$.
Here, $\ket{\cdots}$ denotes the original position basis, whereas $|\cdots)$ denotes symmetric basis states.
Other symmetric states on the $l$th layer of branch $m$ are constructed as [Fig.~\ref{fig:graph} (c)]
\begin{equation}
    |l)_m
    \coloneqq
    \frac{1}{\sqrt{K^{l-1}}}
    \sum_{j=1}^{K^{l-1}} \ket{l,j,m},
    \qquad
    l=1,\dots,M,\quad m=1,\dots,K+1.
\end{equation}
An eigenstate in this symmetric sector is expanded as
$    \ket{\Psi}
    =
    \psi_0 |0)
    +
    \sum_{l=1}^{M}\sum_{m=1}^{K+1} \psi_{l,m} |l)_m .
$
We refer to the eigenstate in this symmetric sector as symmetric eigenstates.
Details of the complementary nonsymmetric basis construction are given in Ref.~\cite{SH-Cayley}.

Within the symmetric sector, the eigenvalue equation reduces to a set of recurrence relations.
In particular, for the class of solutions with $\psi_0=0$, one finds~\footnote{Here we introduce  $\psi_{0,m}\coloneqq \psi_0/\sqrt{K}$ for notational convenience.}
\begin{align}
     E \psi_{l,m}
    &= \sqrt{K}\, t_{\rm R}\, \psi_{l-1,m}
     + \sqrt{K}\, t_{\rm L}\, \psi_{l+1,m}, \\
    \psi_{0,m} &= \psi_{M+1,m}=0, \\
    0 &= \sum_{m=1}^{K+1}\psi_{1,m},
\end{align}
for $l=1,\cdots, M$ and $m=1,\cdots,K+1$.
The first two lines are identical to the eigenvalue problem of the Hatano--Nelson chain, and thus yield
\begin{align}
    \psi_{l,m}^{(n)} = \beta^l \sin(\theta_n l),
    \qquad
    E_n = 2\sqrt{K t_{\rm R} t_{\rm L}}\,\cos\theta_n,
    \label{eq:sol-psi0}
\end{align}
with $\beta=\sqrt{t_{\rm R}/t_{\rm L}}$ and $\theta_n=n\pi/(M+1)$ for $n=1,\dots,M$.
Accordingly, the symmetric eigenstates take the form
\begin{align}
    \ket{\Psi_n}
    =
    \sum_{m=1}^{K+1}\sum_{l=1}^{M}
    c_m\,\psi_{l,m}^{(n)}\,|l)_m,
    \label{eq:sol-sym}
\end{align}
where the coefficients satisfy $\sum_{m=1}^{K+1} c_m = 0$.
This constraint reduces the branch degeneracy from $K+1$ to $K$, so that the symmetric sector contains $KM$ linearly independent eigenstates. 
It should be noted that the obtained symmetric eigenstates are not degenerate with any of the remaining eigenstates if $M+1$ is chosen to be a prime number~\cite{SH-Cayley}.
Since the symmetric solutions are free from $M$-dependent degeneracy at each $n$, the multifractal dimensions are uniquely determined in this sector, making it particularly tractable.

\subsection{Exact multifractal dimension}
\begin{figure}[t]
\centering
\includegraphics[width=\linewidth]{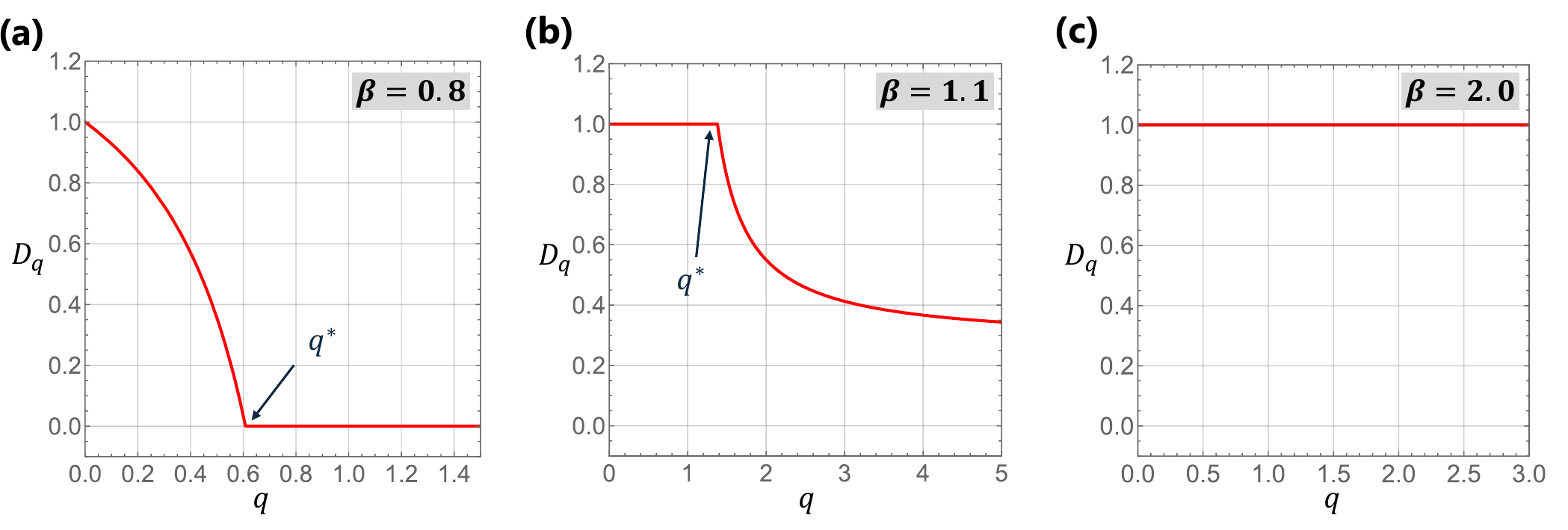} 
\caption{Dependence of the multifractal dimensions $D_q$ on $q$ in the three regimes: (a) $0<\beta<1$, (b) $1<\beta<\sqrt{K}$, and (c) $\beta>\sqrt{K}$. The curves are shown for $K=2$, with representative values $\beta=0.8$, $1.1$, and $2.0$ for panels (a), (b), and (c), respectively. Figure~\ref{fig:Dq} adapted with permission from Ref.~\cite{SH-Cayley}.  \href{https://doi.org/10.1103/PhysRevB.111.075162} {Copyright (2025) by the American Physical Society.}}
\label{fig:Dq}
\end{figure}

We now evaluate the multifractal dimensions analytically.
Substituting the exact symmetric eigenstates in Eq.~\eqref{eq:sol-sym} into the definition of the inverse participation ratio in the position basis $\ket{l,j,m}$, we obtain
\begin{equation}
\label{eq:Iq-sym}
    I_q
    =
    \left[
    \frac{\beta^2 -1}{ \beta^2 (\beta^{2M}-1)}
    \right]^q
    \frac{
    1-\left(\frac{\beta^{2q}}{K^{q-1}}\right)^M
    }{\beta^{-2q}-{K^{1-q}}}
    \sum_{m=1}^{K+1} \abs{c_m}^{2q}.
\end{equation}
Importantly, while the Hilbert-space dimension grows as $\mathcal{N}\propto K^M$ because of the hierarchical branching of the tree, the inverse participation ratio scales either as $I_q \propto \beta^{-2qM}$ or $I_q \propto  {(\beta^{2q}/K^{q-1})}^M$ due to the skin effect.
The competition between geometric branching $K$ and nonreciprocal hopping $\beta$ gives the multifractal properties of the skin modes.
Depending on $\beta$, the multifractal dimensions are classified into three distinct regimes, for which closed-form expressions can be derived analytically.

For $0<\beta<1$, the multifractal dimensions are given by
\begin{equation}\label{eq:Dq1}
    D_q =
    \begin{cases}
        1 - \dfrac{q}{q-1}\dfrac{\log(\beta^2)}{\log K},
        & (q<q^*), \\[4pt]
        0,
        & (q>q^*),
    \end{cases}
\end{equation}
where
\begin{align}
    q^* \coloneqq \frac{\log K}{\log K-\log(\beta^2)}.
\end{align}
Since $D_q$ depends explicitly on $q$, this regime exhibits multifractality [Fig.~\ref{fig:Dq} (a)].
This contrasts with the vanishing multifractal dimensions of single-particle skin modes in one dimension discussed in Sec.~\ref{sec:single}, and reflects the unique scaling behavior of skin modes on a hierarchical lattice.
As $\beta \to 0$, the crossover value $q^*$ tends to zero, and correspondingly $D_q$ vanishes for all $q$.
This limit describes complete localization of the eigenstates around the central node, induced by strong inward nonreciprocity ($t_{\rm L}\gg t_{\rm R}$).

For $1<\beta<\sqrt{K}$, the multifractal dimensions are given by
\begin{equation}\label{eq:Dq2}
    D_q =
    \begin{cases}
        1,
        & (q<q^*), \\[4pt]
        \dfrac{q}{q-1}\dfrac{\log(\beta^2)}{\log K},
        & (q>q^*).
    \end{cases}
\end{equation}
As in the previous regime, the explicit $q$ dependence of $D_q$ signals multifractality [Fig.~\ref{fig:Dq} (b)].
A particularly notable feature appears as $\beta\to\sqrt{K}$, where the crossover value $q^*$ diverges and the multifractal dimensions approach $D_q=1$ for all $q$.
This corresponds to complete delocalization of the eigenstates over the Cayley tree.
Unlike the localized limit $D_q=0$, which is reached only as $\beta\to 0$, full delocalization already occurs at the finite threshold $\beta=\sqrt{K}$.
Physically, even when hopping is biased outward, the wave-function amplitude is redistributed over $K$ descendant branches at each node of the tree. As a result, outward hopping must be strong enough to compensate for this repeated branching. The condition $\beta=\sqrt{K}$ is precisely the point at which these two effects balance, leading to fully delocalized skin modes.
We also note that a finite value of $D_{q \rightarrow \infty} = \log{(\beta^2)}/\log{K}$ means that even the largest wave-function amplitudes are not concentrated on a finite number of sites, but instead scale over a Hilbert-space subset of nonzero effective dimension.

For $\beta > \sqrt{K}$, the multifractal dimensions are
\begin{equation}\label{eq:Dq3}
    D_q = 1,
\end{equation}
so that the eigenstates are fully delocalized in Hilbert space [Fig.~\ref{fig:Dq} (c)].
In the extreme limit $\beta\to\infty$, the wave function is pushed toward the outer boundary of the tree.
However, since the number of boundary nodes increases exponentially with the number of layers and dominates the Hilbert-space volume in the thermodynamic limit, this boundary accumulation is still delocalized from the viewpoint of Hilbert-space occupation.
Accordingly, the states remain characterized by $D_q=1$.

\begin{figure}[h]
\centering
\includegraphics[width=\linewidth]{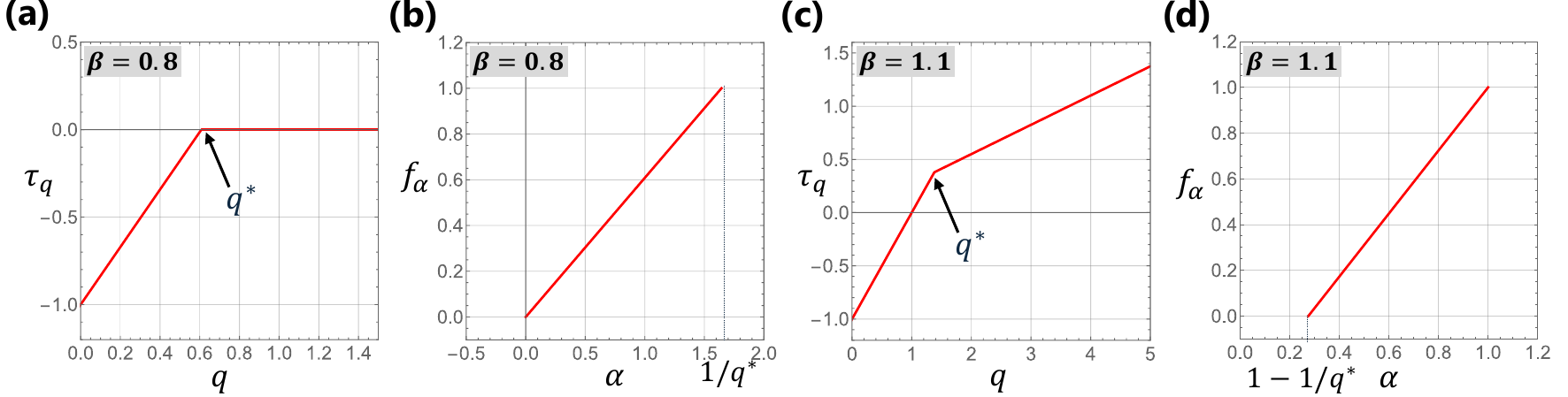} 
\caption{The dependence on $q$ of the exponent
$\tau_q$, and the dependence on $\alpha$
of the multifractal spectrum $f_\alpha$, displayed for $\beta = 0.8$ (a1, a2) and for $\beta = 1.1$ (b1, b2).
The data are
plotted for the case $K=2$. Figure~\ref{fig:fq} adapted with permission from Ref.~\cite{SH-Cayley}. \href{https://doi.org/10.1103/PhysRevB.111.075162} {Copyright (2025) by the American Physical Society.}
}
\label{fig:fq}
\end{figure}

For both $0<\beta<1$ and $1<\beta<\sqrt{K}$, the symmetric eigenstates exhibit multifractal statistics.
The nature of this multifractality, however, differs qualitatively between the two regimes [Fig.~\ref{fig:Dq} (a, b)].
To make this distinction more explicit, we further examine the exponents $\tau_q$ and the corresponding multifractal spectrum $f_\alpha$.
For $0<\beta<1$, we obtain
\begin{align}
\label{eq:a-tauq}
    \tau_q =
    \begin{cases}
        \dfrac{1}{q^*}(q-q^*),
        &(q<q^*), \\[4pt]
        0,
        &(q>q^*),
    \end{cases}
    \qquad
    f_\alpha =
    \begin{cases}
        q^*\alpha,
        &\left(0<\alpha<\dfrac{1}{q^*}\right), \\[4pt]
        -\infty,
        &(\text{otherwise}),
    \end{cases}
\end{align}
whereas for $1<\beta<\sqrt{K}$,
\begin{align}
\label{eq:b-tauq}
    \tau_q =
    \begin{cases}
        q-1,
        &(q<q^*), \\[4pt]
        \dfrac{\log(\beta^2)}{\log K}\,q,
        &(q>q^*),
    \end{cases}
    \qquad
    f_\alpha =
    \begin{cases}
        q^*\alpha + (1-q^*),
        &\left(1-\dfrac{1}{q^*}<\alpha<1\right), \\[4pt]
        -\infty,
        &(\text{otherwise}).
    \end{cases}
\end{align}

For $0<\beta<1$, the exponent $\tau_q$ vanishes for $q>q^*$ [Fig.~\ref{fig:fq} (a)].
This behavior is reminiscent of sparse multifractal states, where the moments at large $q$ are dominated by a small number of exceptionally large amplitudes.
Indeed, in the Anderson transition in high dimensions, one finds $\tau_q=0$ above a finite threshold in the $d\to\infty$ limit~\cite{Mildenberger-02,EFETOV1990119,Fyodorov1992,Mirlin94a,Mirlin94b}.
Equation~\eqref{eq:a-tauq} shows that the skin modes on the Cayley tree exhibit an analogous structure, although the threshold now depends on the parameter $q^*$.
Physically, the moment $I_q$ for large $q$ is controlled by the largest wave-function components, and the condition $\tau_q=0$ therefore implies that the state is strongly concentrated near the central region of the tree, consistent with Eq.~\eqref{eq:sol-psi0}.
By contrast, for smaller $q$, which probe the more broadly distributed part of the wave function, the states retain multifractal character.
The corresponding multifractal spectrum $f_\alpha$ is peaked near $\alpha \simeq 1/q^*$ [Fig.~\ref{fig:fq} (b)], indicating that the wave function scales locally as $|\psi_j|^2 \propto \mathcal{N}^{-1/q^*}$ over a broad set of sites.
We also note that the nonanalyticity of $\tau_q$ at $q=q^*$, together with the saturation $\tau_q=0$ for $q > q^*$, signals a \emph{freezing transition} in the multifractal statistics.
Such frozen behavior has previously been discussed for the Anderson transition on the Bethe lattice~\cite{Luca-14}, the two-dimensional random Dirac model~\cite{Chamon-PRL-96}, and the Rosenzweig-Porter random matrix model~\cite{Kravtsov-15}.
Here, however, the same type of singular multifractal structure arises from nonreciprocity rather than disorder.

For $1<\beta<\sqrt{K}$, by contrast, the exponent $\tau_q$ remains positive for all $q>1$ [Fig.~\ref{fig:fq} (c)].
This is different both from the $0<\beta<1$ regime and from sparse multifractality at high-dimensional Anderson transitions, where $\tau_q$ saturates to zero at large $q$.
The absence of such saturation implies that the wave function does not develop a small set of sharply dominant peaks, even though it remains multifractal.
This is also reflected in the multifractal spectrum $f_\alpha$, which is peaked near $\alpha \simeq 1$ [Fig.~\ref{fig:fq} (d)].
Accordingly, the wave-function amplitudes scale as $|\psi_j|^2 \propto \mathcal{N}^{-1}$ over most of the occupied sites, consistent with the fact that $D_q=1$ for small $q$ in Fig.~\ref{fig:Dq} (b).

\subsection{Multifractal dimensions of degenerate states}\label{sec:weak-disorder}
\begin{figure}[h]
\centering
\includegraphics[width=\linewidth]{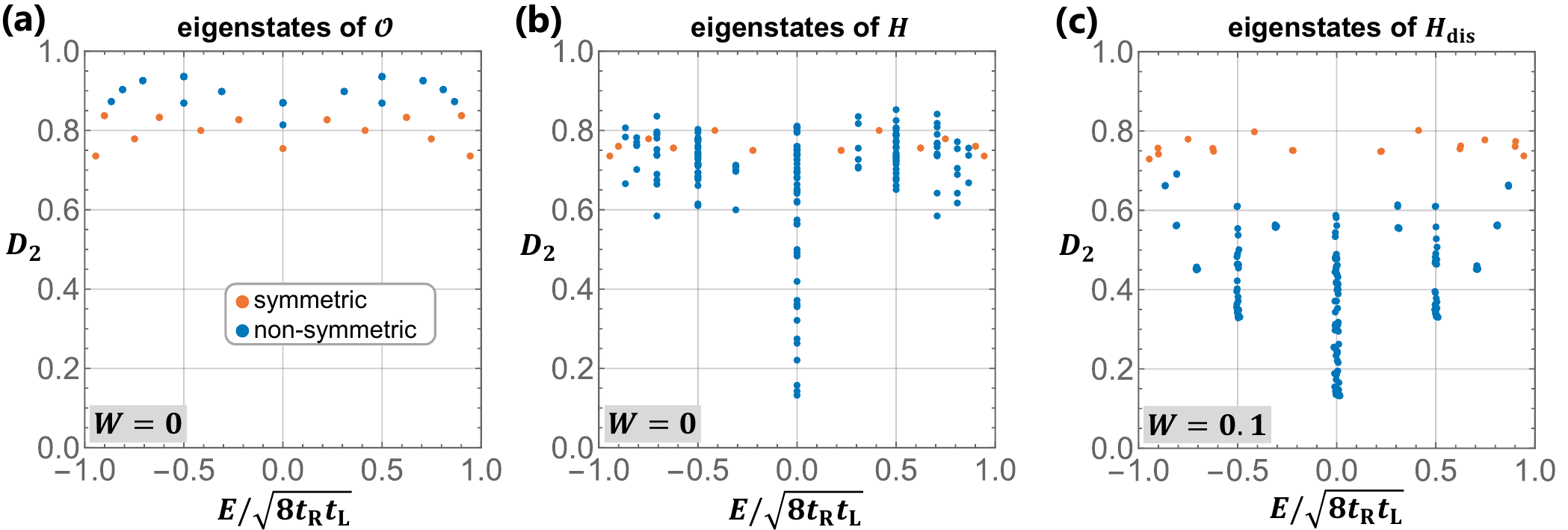} 
\caption{The multifractal dimension $D_2$ as a function of the rescaled energy $E/\sqrt{8 t_{\rm R}t_{\rm L}}$ for a Cayley tree with $K=2$, ($M+1=7, t_{\rm R}=1.1, t_{\rm L}=0.9$). (a, b) In the absence of disorder ($W=0$) and (c) in the presence of weak disorder ($W=0.1$).
Data for symmetric (non-symmetric) eigenstates are shown in orange (blue).
In panel (a), we consider simultaneous eigenstates of the commuting set $\mathcal{O}=\{H, C_3, C_{2,1},\ldots C_{2,M-1}\}$, where $C_3$ denotes the threefold rotation that cyclically permutes the three main branches attached to the central node, and $C_{2,l}$ denotes the twofold swap symmetry at layer $l$, which exchanges the two descendant subbranches branching from each node in that layer.
In panel (b), we show generic eigenstates of the Hamiltonian $H$ without imposing such symmetry-adapted linear combinations.
Figure~\ref{fig:disor} adapted with permission from Ref.~\cite{SH-Cayley}. \href{https://doi.org/10.1103/PhysRevB.111.075162} {Copyright (2025) by the American Physical Society.}
}
\label{fig:disor}
\end{figure}
So far, we have focused on the symmetric eigenstates, for which the multifractal dimensions are uniquely defined.
By contrast, the remaining non-symmetric eigenstates on the Cayley tree are extensively degenerate, and their multifractal dimensions are therefore not uniquely determined.
In this subsection, we show explicitly that the multifractal dimension depends on the choice of linear combination within the degenerate subspace, and we also examine the effect of weak disorder.

For $K=2$, this dependence can be seen by comparing two different constructions of degenerate eigenstates.
When one chooses symmetry-adapted linear combinations that are simultaneous eigenstates of the branch-permutation symmetries of the Cayley tree, the resulting states are more uniformly distributed over the branches and exhibit relatively large values of $D_2$ [Fig.~\ref{fig:disor} (a)].
By contrast, generic linear combinations chosen at random lead to a broader distribution and smaller values of $D_2$, explicitly demonstrating that the multifractal dimension of the degenerate states depends on the particular linear combination chosen [Fig.~\ref{fig:disor} (b)].

We next consider weak disorder, which is expected in realistic experimental settings.
Specifically, we consider
\begin{equation}
H_{\rm dis}=H+\sum_{j=1}^{\mathcal N} U_j \ket{j}\bra{j},
\end{equation}
where $U_j\in[-u,u]$ is a random on-site potential.
In this case, disorder favors linear combinations that are more localized on particular branches, leading to smaller values of $D_2$ for the non-symmetric sector.
The symmetric eigenstates, on the other hand, are not subject to such extensive degeneracy and remain essentially unaffected by weak disorder.
Thus, while the multifractal dimensions of symmetric eigenstates are robust, the highly degenerate non-symmetric eigenstates are strongly sensitive both to the choice of linear combinations and to weak disorder.

\section{Conclusion and outlook}\label{sec:conclusion} 
In this article, we have reviewed the multifractal aspects of the non-Hermitian skin effect across a wide range of settings, from single-particle systems to many-body and tree models.
A central message is that the many-body skin effect gives rise to multifractal structures in many-body Hilbert space, whereas conventional single-particle skin modes on crystalline lattices do not exhibit multifractality. 
We have also highlighted that this multifractality can coexist with random-matrix spectral statistics, unlike conventional multifractal phases such as many-body localized phases. 
Furthermore, the non-Hermitian model on the Cayley tree provides an analytically tractable setting in which the competition between hierarchical branching and nonreciprocity determines the multifractal dimensions exactly.
Taken together, these results provide a unified perspective on multifractal structures associated with the non-Hermitian skin effect. 

In the context of Anderson localization and many-body localization, (multi)fractality has been effectively described by random-matrix models such as power-law banded random matrices~\cite{Mirlin-96} and the Rosenzweig-Porter model~\cite{Rosenzweig-PR-60,Kravtsov-15}, where multifractal dimensions can be obtained analytically.
While several works have explored non-Hermitian extensions of such settings, including the absence of a fractal phase in the non-Hermitian Rosenzweig-Porter model~\cite{De-PRB-22}, much remains to be understood about multifractality in non-Hermitian random-matrix models.

Finally, very recently, nonlinear extensions of the non-Hermitian skin effect have begun to attract increasing attention~\cite{Yuce-21,Ezawa-PRB-22,Yuce-PRB-25,Longhi-25,Kawabata-PRL-25,SH-Koopman}.
The results reviewed here suggest that skin modes are not always most naturally characterized in real space alone, but may instead admit a quantitative description in an appropriate space, such as Hilbert space or graph space, depending on the setting.
From this viewpoint, it would be natural to ask whether nonlinear skin effects may also admit a useful characterization in function space~\cite{SH-Koopman}.
Exploring this possibility remains an important direction for future research.

%
%

\ack{I am deeply grateful to Kohei Kawabata for introducing me to this topic and for teaching me most of the findings in Ref.~\cite{SH-Many}.
I also thank Askar A. Iliasov, Titus Neupert, Tom\'a\v{s} Bzdu\v{s}ek, and Tsuneya Yoshida for related collaborations and valuable suggestions.
I further thank Tsuneya Yoshida for his continued encouragement.
I also thank Jacopo Gliozzi, Taylor L. Hughes, David A. Huse, and Shinsei Ryu for valuable feedback on these works.
Finally, I would like to thank the editors for inviting me to write this paper.}

\funding{This work was supported by JSPS Research Fellow No. 24KJ1445.}





\bibliographystyle{unsrt}
\bibliography{ref.bib}

\end{document}